\UseRawInputEncoding 
\documentclass[journal=jpclcd,manuscript=letter]{achemso}

\usepackage[version=3]{mhchem} 
\usepackage[T1]{fontenc} 
\usepackage{multirow}
\usepackage{amsmath, amssymb}
\usepackage{graphicx}
\usepackage{caption} 
\usepackage{makecell}

\usepackage{color}

\author{Luo Yan}
\affiliation{School of Physics, University of Electronic Science and Technology of China, Chengdu 610054, China.}
\author{Kui xue}
\affiliation{School of Physics, University of Electronic Science and Technology of China, Chengdu 610054, China.}
\author{Jing Zhang}
\affiliation{School of Physics, University of Electronic Science and Technology of China, Chengdu 610054, China.}
\author{Ruiqi Ku}
\affiliation{School of Physics, Harbin Institute of Technology, Harbin 150001, China.}
\author{Bao-Tian Wang}
\affiliation{Institute of High Energy Physics, Chinese Academy of Science (CAS), Beijing 10049, China.}
\author{Liujiang Zhou}
\email{ljzhou86@uestc.edu.cn}
\affiliation{School of Physics, University of Electronic Science and Technology of China, Chengdu 610054, China.}
\alsoaffiliation{Yangtze Delta Region Institute (Huzhou), University of Electronic Science and Technology of China, Huzhou 313001, China.}

\title[An \textsf{achemso} demo]
  {Halogenation induced transition of superconductor-to-semiconductor in MXene-like MOene with direct band gap and long carrier lifetime}

\begin{document}

%
%
%
%

\begin{abstract}
   Traditional MXenes with intriguing mechanical and electronic properties, together with the fertilities of elemental compositions and chemical decorations have aroused much attentions. However, the semiconducting traits with direc band gap are extremetely rare among reported MXenes. Thus, broadening the family of MXene beyond carbides and nitrides with unique behaviors is still an extraordinary and fascinating field. By using the first-principles calculations, the most stable single-layer (SL) dititanium oxide Ti$_{2}$O with edge-sharing trigonal prisms (labled as 1H-Ti$_{2}$O) has been obtained with particle-swarm optimization methods. Similar to the SL 1T-Ti$_{2}$O MOene proposed in our previous work, here, 1H-Ti$_{2}$O monolayer is also an electride material that anionic electron existis on its surface, and an intrinsic Ising superconductor with transition temperature \textit{T}$_{c}$ of $\sim$ 4.7 K. Moreover, the electronic properties of SL 1H- and 1T-Ti$_{2}$O are strongly assocated with the functional groups. In particular, halogenated 1H- and 1T-Ti$_{2}$O monolayers, namely, 1H- and 1T-Ti$_{2}$OX$_{2}$ (X=F, Cl) are semiconductors with direct band gap of 0.58$-$1.18 eV, indicating their attractive performance in optoelectronic and photovoltaic fields. The absorption spectra indicate that SL 1H- and 1T-Ti$_{2}$OX$_{2}$ have strong light harvest ability from ultraviolet to near-infrared region, favoing their potential applications in solar cells and infrared detectors. Furthermore, via the \textit{ab initio} nonadiabatic molecular dynamics, the carrier lifetimes of 1H- and 1T-Ti$_{2}$OF$_{2}$ monolayer are evaluated to be 0.39 and 2.8 ns, respectively, which are comparable to that of MoS$_{2}$ monolayer  (0.39 ns) and multilayer halide perovskites (0.65$-$1.04 ns). Here, the SL 1T-Ti$_{2}$OF$_{2}$ with a larger band gap and a weaker nonradiative coupling exhibits a prolonged electron-hole recombination when compared with 1H-Ti$_{2}$OF$_{2}$ monolayer. This finding further broadens the family of traditional MXenes to MOenes with multifunctional and adjustable properties, and would motivate more efforts in theories and experiments about MOenes family.
\end{abstract}

\section{Introduction}

\section{Introduction}
Since the first discovery of Ti$_{3}$C$_{2}$ monolayer in 2011,\cite{naguib2011two}  the family of two-dimensional (2D) transition metal carbides, nitrides and carbonitrides, collectively called MXenes, have aroused tremendous attentions in energy storage and harvesting,\cite{anasori20172d,yoon2016low} enviromental, electronics,\cite{zhang2019graphene,khazaei2017electronic} catalysts,\cite{peng2019surface,wang2018clay} smart textile, electromagnetic interference shielding and antennas,\cite{shahzad2016electromagnetic} biomedicine,\cite{huang2018two,lin2018insights} and sensors fields \cite{sinha2018mxene,xu2016ultrathin} \textit{etc}.  The first MXene generation is synthesized using a top-down selective etching process with layered ternary MAX phase as precursors.\cite{naguib2013new,naguib201425th,ghidiu2014synthesis} Since then, the continuing diversity in MXenes synthesis approaches have been developed, including HF-forming etching,\cite{naguib2012two,naguib2013new,anasori2015two} molten salt etching,\cite{karlsson2015atomically,li2021halogenated} alkali etching,\cite{li2018fluorine,zou2017heterogeneous} electrochemical etching,\cite{yang2018fluoride, pang2019universal} 
lewis acid molten salt etching,\cite{li2020general} and polar orgainc solvents etching.\cite{natu20202d} In order to prepare monolayer MXenes, organic/inorganic intercalation and delamination \cite{mashtalir2013intercalation,han2019boosting} as well as mechanical delamination\cite{huang2020facile} have also been developed. As a consequence, MXenes have been growing at an unprecedented rate in recent years, on account of the multi chemical ratios, various surface terminations, and formations of carbonitrides and solid solutions.\cite{naguib2021ten} Currently more than 40 different stoichiometric MXenes have been experimentally obtained, and more than 100 MXenes (not considering surface terminations) have been theoretically predicted.\cite{naguib2021ten} The family of MXenes are increasing based on their adjustable structure and rich surface chemistry.

Functional groups usually saturate the nonbonding valance electrons of MXene via their low-energy orbitals during the etching process, resulting in significant effects on their physical and chemical characteristics. Typically, non-terminated MXenes are metallic, whereas terminated MXenes behave highly conductive metallic,\cite{naguib2011two,zhang2017transparent,dillon2016highly} half-metallic,\cite{zhou2018proximity,zhang2017robust} semiconducting,\cite{khazaei2013novel,zha2016thermal,jiang2020two,wong2018enhancing,zha2015role,hong2016first} topologically insulating and highly insulating states \cite{fashandi2015dirac,si2016large,khazaei2016topological,weng2015large} depended on the surface terminations. However, as far as we known, only Mo$_{2}$CT$_{x}$ \cite{halim2016synthesis} and Mo$_{2}$TiC$_{2}$T$_{x}$\cite{anasori2016control} show semiconducting  features determined in experiments, and M$_{2}$CO$_{2}$ (M = Ti, Zr, Hf, Mn, and W), M$_{2}$CF$_{2}$ (Cr, and Mo), M$_{2}$CT$_{2}$ (M = Sc, Y; T = O, F, OH) and M'$_{2(1-x)}$M''$_{2x}$CO$_{2}$ (M' and M'' are Ti, Zr, and Hf) solid-solution MXenes exhibit semiconducting behaviors predicted in theories.\cite{khazaei2013novel,zha2016thermal,jiang2020two,wong2018enhancing,zha2015role,hong2016first} More particularly, Sc$_{2}$C(OH)$_{2}$, \cite{lee2015achieving} Y$_{2}$C(OH)$_{2}$,\cite{Wang2019} Cr$_{2}$TiC$_{2}$(OH)$_{2}$\cite{yang2016tunable} are the only three instrinsic direct semiconductors with band gap values of 0.71, 0.72 and 0.84 eV (HSE06 level), respectively. Therefore, searching new MXene materials with direct band gaps is significant for semiconductor field.

Inspiringly, oxygen can substitute for either C or N in transition metals carbides/nitrides within their bulk phases, forming oxycarbides or oxynitrides. \cite{vahidmohammadi2021world} Thus, such substitution in MXenes should be feasible, providing a better understanding of the arrangement of X site atoms in MXenes. Recently, single-layer (SL) Ti$_{2}$B, labled as MBene, has been proposed with metallic behavior and ferromagnetic nature with Curie temperature around 39.06 K,\cite{ozdemir2019new} favoring the crystal diversity within MXenes framework. None the less, the semiconductors have never been reported within MBenes. In contrast, SL Tl$_{2}$O exfoliated from layered bulk phase exhibits a direct band gap of 1.56 eV (HSE06+SOC level) and a very high charge carrier mobility.\cite{ma2017single} However, Tl is generally toxic and scarce. Considering the nontoxic and environmentally friendly Ti-based MXenes, built of abundant Ti elements are particularly attracting considerable attentions. In this regard, SL 1T-Ti$_{2}$O based on transition-metal oxides, namely MOene, have been reported to be electride, anode materials, and superconductor. Intriguingly, fluorinated 1T-Ti$_{2}$O monolayer (1T-Ti$_{2}$OF$_{2}$) is a semiconductor trait with a direct band gap of 1.18 eV (HSE06 level). \cite{yan2020single} However, the 1T phase of MOene has not a global minimum of energy within the stoichiometry of Ti:O = 2:1. On the one hand, the transition of superconductor-to-semiconductor induced by halogenated surface terminations in MOene displays great potentialities in optoelectronic fields. On the other band, the excitonic effects in low dimensional systems and the carrier lifetime in optoelectronic devices are fundamental.

In this work, we obtain another crystal of 1H-Ti$_{2}$O monolayer, using the particle-swarm optimization methods. SL 1H-Ti$_{2}$O is the global minimum one within the stoichiometry of Ti:O = 2:1, and its formation energy is lower by 96.67 meV/atom than 1T phase. The same as 1T-Ti$_{2}$O monolyer, SL 1H-Ti$_{2}$O is also an electride material and superconductor. The electronic properties of functionalized SL 1H and 1T-Ti$_{2}$O are strongly associated with the surface terminations. Surface fluoridation the key to tune the metallic trait of SL 1T-Ti$_{2}$O to a semiconducting state.  Specifically, halogenated 1H and 1T-Ti$_{2}$OX$_{2}$ (X=F, Cl and Br) monolayers are all semiconductors with band gaps ranging from 1.18 to 0.58 eV. Among them, fluorinated and chlorinated 1H and 1T-Ti$_{2}$O are direct semiconductors,  while brominated ones are indirect. When considered exciton effects, their light-absorbing are from ultraviolet to near-infrared region with strong light harvest ability, along with considerably small binding energies. The nonadiabatic molecular dynamics (NAMD) simulations reveal that the carrier lifetimes of SL 1H and 1T-Ti$_{2}$OF$_{2}$ reach to nanosecond level, favoring their great prefermance in optoelectronic devices.

\section{Computational Methods}
\par The first-principle calculations in the framework of density functional theory (DFT) were performed via the Vienna ab initio simulation package (VASP).\cite{kresse1996efficient,kresse1999ultrasoft} The Perdew-Burke-Ernzerhof (PBE) pseudopentials with 500 eV plane-wave energy cutoff and  generalized gradient approximation (GGA) \cite{blochl1994projector,blochl1994improved}  were adopted for the exchange correlations. The existence of the 1T-Ti$_{2}$O  monolayer \cite{yan2020single} was firstly confirmed in our former work, and here other possible 2D structures of the same stoichiometry (Ti:O = 2:1) were systematically searched with CALYPSO code. \cite{Wang2012CALYPSO} The opt88-vdw method was used to model the van der Waals interaction. \cite{klimes2011van} For Brillouin zone (BZ) sampling, 25 $\times$ 25 $\times$ 1 k-point mesh was employed.  In order to avoid the coupling between adjacent layers, the vacuum space was set be 15 \AA{}  along the \emph{z} direction. The phonon properties were carried out using density functional perturbation theory (DFPT) as implemented in PHONOPY code.\cite{togo2015first} The hybrid functional (HSE06)  mixed with 25\% exact Hartree-Fock (HF) exchange was used to correct PBE band gaps. Some data post-processing for VASP calculations were dealed with the help of VASPKIT code.\cite{VASPKIT} Moreover, as implemented in YAMBO software,\cite{marini2009yambo} the screened Coulomb interaction (G$_{0}$W$_{0}$ approximation) in combination with the random phase approximation (RPA) or Bethe-Salpeter equation (BSE) was adopted to calculate the quasi-particle band gap and the light absorbance with or without electron-hole (e-h) interactions. Besides, the carrier lifetimes were evaluated by the \textit{ab initio} NAMD simulations with Hefei-NAMD code. \cite{zheng2019ab} The methods and computational details were avaiable in Supporting Information.

\section{Results and discussion}

Sigle-layer (SL) 1H-Ti$_{2}$O is exhibited in Figures 1a, which is the most stablest one with stoichiometric ratio of Ti:O = 2:1 within particle-swarm optimization framework. 1H-Ti$_{2}$O is in a trigonal prismatic arrangement with an inverted hexagonal symmetry, and crystallizes in hexagonal space group, \emph{P$\bar{6}$m2} (no. 187). To determied the ground state of 1H-Ti$_{2}$O monolayer, possible nonmagnetic (NM), ferromagnetic (FM) and antiferromagnetic (AFM) (i.e., AFM-Néel, AFM-Stripy and AFM-Zigzag) spin configurations are also considered within its conventional cell in this work (Figure S1). Calculated results are presented in Table 1. Similar to 1T-Ti$_{2}$O monolayer (Figure 1b), \cite{yan2020single}  SL 1H-Ti$_{2}$O with AFM-Néel spin configuration is also the most stable state, and its total energy is 96.67 meV/atom lower when compared with SL 1T.  Meanwhile, due to the dangling 3\textit{d} orbitals of Ti atoms, there are 0.59 $\mu$B magnetic moment upon per Ti atom. The optimized lattice parameters are a = 2.80 {\AA}, b = 4.84 {\AA}, and the bond length of Ti$-$O is 2.15 {\AA}, slightly larger than that of SL 1T-Ti$_{2}$O (2.09 {\AA}). Furthermore, Ti$-$O bonds are ionic bonding interactions confirmed by electron localization function (ELF) (Figure S2a), and via which the charge transfer are from Ti to more electronegative O atoms (Figures S2b,c). Besides, the thickness between upper and lower plane of Ti atoms is 2.85 {\AA}, also just slightly higher than 1T-Ti$_{2}$O monolayer (2.82 {\AA}).\cite{yan2020single} The stabilities of 1H-Ti$_{2}$O monolayer are discussed in Supporting Information. Fortunately, bulk Ti$_{2}$O with hexagonal crystal have been established through neutron diffraction studies of the suboxides of titanium in 1970.\cite{kornilov1970neutron} In recent, 1T-Ti$_{2}$O particles with high conductivity have been prepared by a molten salt electrochemical synthesis process. \cite{wang2020controllable} The metallic films of 1T-Ti$_{2}$O have been successfully grown on commercial (0001)-oriented $\alpha$-Al$_{2}$O$_{3}$ single crystalline substrates.\cite{fan2019structure} Therefore, 1T-Ti$_{2}$O monolayer may be achieved via the similar deposition process or delaminating the films of 1T-Ti$_{2}$O uopn the freezing water molecules as observed in MXene.\cite{huang2020facile}

The same rules as Y$_{2}$C and 1T-Ti$_{2}$O monolayer,\cite{park2017strong, yan2020single} as for SL 1H-Ti$_{2}$O, two Ti atoms jointly contribute four valence electrons, while the oxygen anion only has the inclination to accept two. Therefore, the chemical formula of SL 1H-Ti$_{2}$O also can be expressed as [Ti$_{2}$O]$^{2+}$$\cdot$2e$^{-}$. Subsequently, SL 1H-Ti$_{2}$O is also confirmed to be an electrde meterial with anionic electrons above the center of Ti atoms (Figure S2a). Here, the electride states originating from the surplus electron serve as anionic electron in geometric space upon the Ti atoms' interreactions. Meanwhile, the difference charge plots show the charge are transferred from less-electronegative Ti to more electronegative O atoms and anionic electron site (labeled as X1 and X2 pseudo atoms in Figures S2d, e. The calculated spin-up band structure with projected electride states is shown in Figure 2a. Similar to the previously reported Y$_{2}$C\cite{park2017strong}, narrower bands consisted of electron and hole pockets cross the Fermi level, indicating its metallic feature. From the partial density of states (PDOS), the metal nature is mostly contribute by the Ti-3\textit{d} orbitals, especially for Ti-\textit{d}$_{x^{2}-y^{2},xy}$ and Ti-\textit{d}$_{xz,yz}$ orbitals (Figure S4a). Besides, the anionic electron states (X-band) are mostly in the range of $-$3 eV < E$-$E$_{F}$ < 0 eV and occupied, which is in line with 1T-Ti$_{2}$O monolayer, \cite{yan2020single} indicating a large amount of anionic electrons in SL Ti$_{2}$O crystals. Clearly, the anionic electron have significant hybridization with Ti-3\textit{d} orbitals, thus, the electride states mainly originate from unsaturated Ti-3\textit{d} orbitals.\cite{yan2020single} Furthermore, the electronic workfunction of 1H- and 1T-Ti$_{2}$O are calculated to be 4.55 and 4.87 eV, respectively (Figire S5), comparable to mostly reported bare MXene (4$-$5 eV). \cite{khazaei2015oh} The same as 1T-Ti$_{2}$O monolayer,\cite{yan2020single} SL 1H-Ti$_{2}$O is also a 2D superconducor with the superconducting transition temperature \textit{T$_{c}$} of 4.7 K upon $\mu$$^{*}$ = 0.1 (Figures S4b-d). More detailed discussions about its superconductivity are available in Supporting Information.

The exposure of the chemically active outer layer transition metals of MOene would integrate with the surface terminations easily. Here, the  surface functional groups, such as typical hydrogenation, hydroxylation, oxidization and fluoridation group, are further studied on 1H-Ti$_{2}$O monolayer. Three major possible configurations of functionlized termination modes are considered (Figures S6, 7) and the structural stabilities of different functionalized derivatives can be estimated by comparing their total energies. To avoid the steric repulsion from the underneath Ti atoms, the  the most stable sites for fluoridation and hydroxylation groups is above the centers of hexatomic ring of Ti atoms (Figures S6c and S7c). Besides, the hydrogenation and oxidization groups loacte above the centers of hexatomic ring of Ti and O atoms(Figures S6a and S7a). Importantly, the surface passivations saturate the dangling Ti-3\textit{d} orbitals, resulting in their nonmagnetic ground states. Henceforth, their stability and electron properties were explored within their primitive cells. Phonon spectra analyses reveal that 1H-Ti$_{2}$OH$_{2}$, Ti$_{2}$O(O)$_{2}$ Ti$_{2}$O(OH)$_{2}$ and Ti$_{2}$OF$_{2}$ are all kinetic stability (Figure S8). Compared with unstable SL 1T-Ti$_{2}$O(O)$_{2}$,\cite{yan2020single} 1H-Ti$_{2}$O monolayer with lower energy shows higher stability under ambient conditions. The lattice constants of SL 1H-Ti$_{2}$OH$_{2}$, Ti$_{2}$O(O)$_{2}$, Ti$_{2}$O(OH)$_{2}$ and Ti$_{2}$OF$_{2}$ are optimized to be a $=$ b $=$ 2.80, 3.10, 2.88 and  2.83 {\AA}, respectively, comparable to those of functionalized  1T-Ti$_{2}$O monolayers (2.83$-$2.89 {\AA}).\cite{yan2020single} The electronic properties of Ti$_{2}$O layers are strongly associated with the surface terminations. SL 1H-Ti$_{2}$OH$_{2}$ and Ti$_{2}$O(OH)$_{2}$ show semimetallic features (Figures S9a,b), which is similar to 1T-Ti$_{2}$OH$_{2}$ and Ti$_{2}$O(OH)$_{2}$ \cite{yan2020single} monolayer, serving as anode material candidates.\cite{tang2012mxenes} In contrast, SL 1H-Ti$_{2}$O(O)$_{2}$ is an indirect semiconductor (band gap of 1.5 eV in Figure S9c), and Ti$_{2}$OF$_{2}$ is a direct semiconductor (band gap of 0.82 eV in Figure S9d), due to electrons transfer from the transition metal to strong electronegative surface terminations. Besides, the electronic workfunctions in functionalized 1H- and 1T-Ti$_{2}$O  can be shifted in the range of 2.47$-$5.40 eV relying on functional groups (Figure S10). Intriguingly, the workfunctions of MOene are not only comparable to prinstine MXenes (3.3$-$4.8 eV)  and the functionalized ones by F (3.1$-$5.8 eV), OH (1.6$-$2.8 eV), and O group (3.3$-$6.7 eV),  but also are accordance with the trend in MXene, following O- > F- > bare > OH- function groups. \cite{khazaei2015oh} In particular, 1H- and 1T-Ti$_{2}$O(OH)$_{2}$ monolayers display the lowest workfunction values of 2.47 and 2.55 eV, respectively, promising for electrical contact, field emitter cathodes and thermionic devices.\cite{jiang2020two}

In partical, as explored by former efforts,  SL 1H- and 1T-Ti$_{2}$OF$_{2}$ are excellent semeconductors with direct band gap values of $\sim$1 eV, presenting an interesting potential application for visible-light solar harvesting/utilizing techniques and good candidates as infrared photodetectors.\cite{yu2019two} For completeness, we systematically study the halogenated 1H- and 1T-Ti$_{2}$O monolayer, namely 1H- and 1T-Ti$_{2}$OX$_{2}$ (X=F, Cl, Br and I) (Figure 1). As shown in Figure S11, phonon spectra analyses confirm the dynamic stability of the fluorinated, chlorinated and bromized 1H- and 1T-Ti$_{2}$O. However, there are some imaginary modes of 1H- and 1T-Ti$_{2}$OI$_{2}$ in the entire Brillouin zone. Therefore, we next only focu on SL 1H- and 1T-Ti$_{2}$OX$_{2}$ (X = F, Cl and Br). As expected, 1H- and 1T-Ti$_{2}$OX$_{2}$ monolayer are without any symmetry constraints, and also belong to the \emph{P$\bar{6}$m2} and \emph{P$\bar{3}$m1} space group, respectively. Some relaxed structural parameters are listed in Table S1. Obviously, the lattice constans of SL 1H- and 1T-Ti$_{2}$OX$_{2}$ (X = F, Cl and Br) increase steadily in halogen group monotonically. Compared with printine 1H- and 1T-Ti$_{2}$O monolayer, the equilibrium lattice parameters of 1H- and 1T-Ti$_{2}$OBr$_{2}$ are enlarged about 13.1\%, due to the increase of atomic radius of X atoms. The bond length have a similar trend with equilibrium lattice parameters. This is understandable because the electronegativity of X atom decreases in halogen group, leading to weaker bonding interactions between Ti$-$O and Ti$-$X. Thus, their thicknesses are also proportional with the bond lengthes and elongates going down. As a consequence, their total energies (E$_{tot}$) decrease in halogen group (Table S1). In general, fluorinated 1H phase of Ti$_{2}$OF$_{2}$ is more stable than 1T type. However, when chlorinated and bromized, 1T-Ti$_{2}$Cl$_{2}$ and 1T-Ti$_{2}$Br$_{2}$ monolayers are more energetically stable than 1H ones, which coincides with following thermal stability simulations. Moreover, the stabilities of Fluorinated SL 1H- and 1T-Ti$_{2}$O are discussed in Supporting Information.

The projected band structures (Pband) and PDOS split into the Ti, O and X contributions show that valence and conduction band of SL Ti$_{2}$OX$_{2}$ are essentially dominated by the Ti-3$\textit{d}$ orbitals, especially for finite Ti-\textit{d}$_{x^{2}-y^{2},xy}$ states (Figure S13). As indicated in Figures 2b, c, SL 1H- and 1T-Ti$_{2}$OF$_{2}$ are direct semiconductors with band gap values of 0.82 and 1.18 eV, respectively. Their valence band maximum (VBM) and conduction band minimum (CBM) are located at the $\Gamma$ point, leading to a directly allowed electronic transition between them, which is confirmed by the sum of squares of the transition dipole moment (P$^{2}$). The parities of VBM and CBM at $\Gamma$ point are also opposite, further favoring a unforbidden electronic transition between them. According to the Bader calculations,\cite{tang2009grid} as for 1H-Ti$_{2}$OF$_{2}$ monolayer, the net charger transfer from Ti to O and F are 1.29 and 0.71 \textit{e} per atom, respectively, which can be represented as 1H-Ti$_{2}$$^{1.36-}$O$^{1.29+}$F$_{2}$$^{0.71+}$. The ionic feature of 1T-Ti$_{2}$OF$_{2}$ also can be elucidated as 1T-Ti$_{2}$$^{1.4-}$O$^{1.37+}$F$_{2}$$^{0.71+}$, implying charge transfer entirely from Ti to more electronegative O and F atoms. In case of surface chlorination, SL 1H- and 1T-Ti$_{2}$OCl$_{2}$ are also direct-band-gap semiconductors with VBM and CBM locating at the $\Gamma$ point (Figures S14a, d), along with  allowed electronic transitions (Figures S14b, e). Compared with fluorinated Ti$_{2}$O, the band gap values slightly decrease to be 0.58 and 1.12 eV for 1H- and 1T-Ti$_{2}$OCl$_{2}$ monolayer. Due  to the decrease of electronegativity, the net charge transfering from Ti to Cl atom decreases to be 0.59 and 0.61 $\textit{e}$ per atom among SL 1H- and 1T-Ti$_{2}$OCl$_{2}$. With surface bromination, 1H- and 1T-Ti$_{2}$OBr$_{2}$ monolayer are indirect-band-gap semiconductors with VBM and CBM locating at $\Gamma$ point and K (H) point, respectively (Figures S14g, j).  And their indirect band gap are about 0.07 and 0.92 eV, and direct band gap  are  0.63 and 1.17 eV. P$^{2}$ reveals that the high transition feasibility is available between VBM and VBM at at $\Gamma$ point, rather than from the $\Gamma$ to K point (Figures S14h, k).  Besides, the net charger transfer from Ti to O and Br atom and can be expressed as 1H-Ti$_{2}$$^{1.22-}$O$^{1.34+}$Br$_{2}$$^{0.55+}$ (1H-Ti$_{2}$$^{1.25-}$O$^{1.40+}$Br$_{2}$$^{0.55+}$). In addition, the workfunction of 1H- and 1T-Ti$_{2}$OX$_{2}$ are evaluated (Figure S15) and listed in Table S1. Obviously, their values decrease down the halogen group and vice versa, and are comparable to fluorinated MXene (3.1$-$5.8 eV), \cite{khazaei2015oh} which can be experimentally measured by Kelvin Prove atomic force microscope (KPFM) under air ambient,\cite{mariano2016solution} ultraviolet photoelectron spectroscopy (UPS) under vacuum\cite {wang2018oxide} and photoelectron spectroscopy in air (PESA).\cite{Kang2017}

We next crossover our explorations from electronic aspects to optical properties of 1H- and 1T-Ti$_{2}$OX$_{2}$ (X=F, Cl and Br) monolayers. Generally, large exciton binding energiy arises from the reduced screening in thin monolayer. \cite{zhou2017computational,zhou2018interlayer} Therefore, the accurate optical absorption and excitonic nature are systematically scrutinized based on GW plus random phase approximation (GW+RPA) (without e-h interaction) and Bethe-Salpeter
equation (GW+BSE) (with e-h interaction). \cite{rohlfing2000electron,deslippe2012berkeleygw} The imaginary part of the transverse dielectric constant overlapped with the incident AM1.5G solar flux for 1H- and 1T-Ti$_{2}$OX$_{2}$ monolayers are  shown in Figure 3 and Figure S14. The first prominent peak with e-h interaction corresponds to the first bright excitonic energies (defined as the optical band gap, E$_{g}$$^{o}$). The G$_{0}$W$_{0}$ gap is obtained from the absorption edge within GW+RPA level. Hence, the exciton binding energy (E$_{b}$) can be calculated by the difference between G$_{0}$W$_{0}$ gap and optical band gap. Related results are listed in Table 2.  As  for 1H- and 1T-Ti$_{2}$OF$_{2}$ monolayer (Figures 3a, c), the quasi-particle energy gaps are 0.72 and 1.27 eV, and optical band gaps are 0.28 and 0.76 eV, respectively.  Compared with HSE06 band gap, the GW corrections are about 0.1 and 0.09 eV for SL 1H- and 1T-Ti$_{2}$OF$_{2}$, respectively, revealing their weak excitonic effect. Then, the E$_{b}$ can calculate to be 0.44 and 0.51 eV, which is smaller than that of Janus-MoSTe (0.54 eV), MoS$_{2}$ (0.80 eV), \cite{jin2018prediction} TiNCl (0.74 eV) and TiNBr (0.66 eV) monolayers. \cite{zhou2017computational} Clearly presented in Table 2, the E$_{g}$, G$_{0}$W$_{0}$ gap and E$_{b}$ of SL Ti$_{2}$OX$_{2}$ (X=F, Cl and Br) decrease down the halogen group. To understand the nature of the first peak excitons, the electron-hole wave functions of SL 1H- and 1T-Ti$_{2}$OF$_{2}$ are shown in Figures 3b, d. Clearly, the higher-enery exciton in SL 1T-Ti$_{2}$OF$_{2}$ is more spatially localized than 1H-Ti$_{2}$OF$_{2}$ monolayer, as expected by its higher E$_{b}$. Due to the less ionic behavior of Cl/Br atoms with respect to F atoms, thus, the electronic screening is enhanced and E$_{b}$ is reduced for Ti$_{2}$OCl(Br)$_{2}$ monolayers. Consequently, the bound exciton could distribute more broadly when compared with Ti$_{2}$OF$_{2}$ monolayer. In particular, such considerably small values of E$_{b}$ indicate the excitons among 1H- and 1T-Ti$_{2}$OX$_{2}$ monolayer are relatively readily to be dissociated into free electrons and holes. Intriguingly, their E$_{g}$$^{o}$ are in the range of 0.13$-$0.76 eV, indicating SL Ti$_{2}$OX$_{2}$ are suitable for infrared detectors with strong infrared light absorption. Moreover, according to a large overlap with AM1.5G solar flux, they also have remarkably major absorption in visible region. Thus, Ti$_{2}$OX$_{2}$ monolayers can be applied as donor materials in excitonic solar cells, if paired with suitable acceptors with matching band alignment.

The lifetime of excited carriers (recombination of excited electron and hole) limits material quality in light-to-current conversion efficiency in solar energy applications and nanoscale optoelectronic devices. And the energy of excited carriers can transform to either photons (radiative process) or phonons (nonradiative process). But the nonradiative (NA) charge recombination is the dominating factor due to the too fast radiative process. \cite{abakumov1991nonradiative} Therefore, taking 1H- and 1T-Ti$_{2}$OF$_{2}$ monolayer as an illustration, the NA e-h recombinations were adopted to evaluate the charge carrier lifetimes. The magnitude of the nonradiative coupling (NAC) between initial and final states are crucial for NA e-h recombination rate. Here, particularly since radiative relaxation in semiconductors occurs on subpicosecond time scales, \cite{zhang2018rapid,long2012photo,long2014instantaneous} we presumed that the hot electron and hole have relaxed to CBM and VBM before recombination. Figures 4a, b shows the VBM and CBM edge states of SL 1H- and 1T-Ti$_{2}$OF$_{2}$.  The real-space partial charge density of SL 1H-Ti$_{2}$OF$_{2}$ shows delocalized and continuous features: charge density of the VBM mainly spread over in-plane of Ti atoms, while the charge density of CBM is spread over out-of-plane Ti atoms. The PDOS indicates that valence bands have a strong hybridization within Ti-\textit{d}$_{x^{2}-y^{2},xy}$ states, and the CBM  mainly orginates from Ti-\textit{d}$_{x^{2}-y^{2},xy}$, Ti-\textit{d}$_{z^{2}}$ and O-\textit{p}$_{xy}$ orbitals (Figure S13a). In contrary, 1H-Ti$_{2}$OF$_{2}$ shows stronger localization for CBM and VBM  edge states. The charge density of VBM in SL 1T-Ti$_{2}$OF$_{2}$ locates on the lower-plane of Ti atoms, while the CBM distributes on the upper-plane Ti atoms, in which $\sigma$ and $\pi$ electron are fully separated (Figure S13d). Therefore, the spatially resolved electron and hole wave functions results in a smaller NAC for SL 1T-Ti$_{2}$OF$_{2}$ (2.46 meV) vs 1H-Ti$_{2}$OF$_{2}$(3.10 meV). Thus, both a larger band gap and smaller NAC hint that the NA e-h recombination should be slower in 1T-Ti$_{2}$OF$_{2}$ monolayer.

To clearly characterize the phonon modes that couple to the electronic transition and participation in the NA e-h recombination, the Fourier transforms (FTs) of autocorrelation functions (ACF) of fluctuations of  VBM$-$CBM energy gaps along the MD trajectories are further explored. The spectral densities or influence spectra calculated by FTs of the ACF, are plotted in Figure 4c. The strength of electron-phonon (e-ph) interaction can be reflected by the participation of phonon modes and their corresponding magnitude. The frequency vibrations of Ti$_{2}$OF$_{2}$ in range of 200$-$300 cm$^{-1}$ majorly consisted of out-of-plane Ti and F atoms vibrations are engaged in the NA process.  However, O atoms dominating at higher frequency contribute little to the spectrum density, revealing that higher frequency modes create negligible influence on their NAC (Figures S11a, e).  Unlike other systems, higher frequency modes create larger NAC at a given temperature, due to the NAC is proportional to nuclear velocity. The visualizations of the vibration modes involved in e-ph interaction at $\Gamma$ points are visiblle for 1H- and 1T-Ti$_{2}$OF$_{2}$ monolayer (Figure S16). Moreover, the presence of solitary frequency in the vibrational influence spectrum would decelerate decoherence in Ti$_{2}$OF$_{2}$ monolayer, contrary to halide perovskite. \cite{zhang2018rapid, li2018time} Obviously, the strength of the e-ph interactions in 1H-Ti$_{2}$OF$_{2}$ monolayer is stronger than that of 1T-Ti$_{2}$OF$_{2}$, rationalizing the larger NAC with larger VBM$-$CBM overlap.

The carrier lifetime is determined by the NAC and phonon-induced coherence time (pure dephasing time). Pure-dephasing process indicates elastic electron-phonon (e-ph) scattering, which is much faster than the e-h recombination,\cite{guo2016electron} requiring inclusion of decoherence into NAMD.\cite{habenicht2008nonradiative} The decoherence time, also termed as pure-dephasing time, can be evaluated by the second-order cumulant approximation of optical response theory.\cite{prezhdo1998relationship} On that account, the pure-dephasing times ($\tau$) can be acquired by Gaussian fitting of the pure-dephasing functions, exp[-0.5(t/$\tau$)$^{2}$], exhibited in Figure S17a. The pure-dephasing/decoherence time for SL 1T-Ti$_{2}$OF$_{2}$ (20.11 fs) is relatively longer than SL 1H-Ti$_{2}$OF$_{2}$ (11.26 fs), listed in Table 5. It is usual that the homogeneous line width (HLW) can be obtained by the reduced Planck’s constant divided by the pure-dephasing time. \cite{zhang2018rapid} Thus, the longer pure-dephasing time is corresponding to sharper optical lines in 1T-Ti$_{2}$OF$_{2}$ monolayer. Moreover, the initial amplitude and decay of the ACF is revelent to decoherence, \cite{madrid2009phonon} revealing that a greater initial value and a more asymmetric ACF indicates faster dephasing process. Generally, the square root of the initial value of the  unnormalized ACF (uACF) presents the magnitude of the phonon-induced VBM$-$CBM energy gap fluctuation, depicting inhomogeneous linebroadening. The gap fluctuations are  85.32 and 53.39 meV for SL 1H- and 1T-Ti$_{2}$OF$_{2}$, respectively. Thus, the homogeneous line widths of them shown in Table 5 are little smaller to inhomogeneous contributions. Apparently shown in Figure S17b, the initial uACF value is larger in 1H-Ti$_{2}$OF$_{2}$, rationalizing the faster pure-dephasing in its gap fluctuations. And the large fluctuation corresponds to larger signal magnitudes in the spectral density (Figure 4c), which also supports the stronger e-ph interactions in 1H-Ti$_{2}$OF$_{2}$ monolayer. A periodic oscillatory behavior for uACF also favors that few modes couple to their electron subsystem in e-ph interactions, which is advantageous to decrease of NAC.

The evolutions of the excited electron state population for 1H- and 1T-Ti$_{2}$OF$_{2}$ monolayer with NA e-h recombination were depicted in Figures 4d, e. Before the NA e-h recombination simulations, one electron is excited from the VBM to the CBM. Here, their band gaps were scaled uopn quasi-particle correction as summarized in Table 2. By fitting the data to short-time linear expansion of exponential decay, \textit{f(t)}=exp($-$\textit{t}/$\tau$) $\approx$ 1$-$\textit{t}/$\tau$, the carrier lifetime can be evaluated and summarized in Table 5. SL 1T-Ti$_{2}$OF$_{2}$ has a longer electron relaxation time as e-h recombination time reaches the nanosecond level, suggesting the photoexcited electrons are easier to be collected to form photocurrent. In the framework of Marcus theory \cite{marcus1985electron} and Fermi's golden rule,\cite{hyeon2009symmetric}  a larger band gap, a weaker NAC, and a shorter pure-dephasing time lead to a smaller recombination rate. Here, spatially distinguished conduction and valence band edges of  1T-Ti$_{2}$OF$_{2}$ monolayer reduce the NAC, leading to a prolonged lifetime of the photoexcited charge carriers. Therefore, their instrinsic band gap, NAC magnitude and pure-dephasing time have analyzed in detail above, accounting for the difference in the charge recombination rates. The simulated carrier lifetimes posses the same order of magnitude as former reported systems, such as pristine and doped BP (0.39$-$5.34 ns), \cite{guo2020tuning} MoS$_{2}$ (0.39 ns),\cite{li2017sulfur} Janus-MoSTe monolayer (1.31 ns), \cite{jin2018prediction} multiLayer 2D halide perovskites (0.65$-$1.04 ns)\cite{zhang2018rapid} and MAPbI$_{3}$ perovskite (12.0 ns),\cite{li2018time} favoring their fascinating properties in optoelectronic and photovoltaic fields.

\section{Conclusions}
In conclusion, within particle-swarm optimization framework, we have obatined the most stable SL 1H-Ti$_{2}$O MOene with stoichiometic ratio of Ti:O = 2:1. Similar to early reported 1T-Ti$_{2}$O monolayer, \cite{yan2020single} here, SL 1H-Ti$_{2}$O with reliable stability also is also an  electride material with anionic electron residing on the surface of Ti atoms, and an instrinsic 2D superconductor with \emph{T$_{c}$} of $~$ 4.7 K. Moreover, the electronic properties of Ti$_{2}$O monolayers are strongly related to the surfacee terminations. The hydrogenated and hydroxylated SL Ti$_{2}$O show semimetallic features, serving as anode materials candidates, which can be verifed by further  studies.  While Ti$_{2}$O monolayers turn into semiconductors when oxidized and fluorinated. Specifically, fluorinated SL 1H- and 1T-Ti$_{2}$O are fantastic seneconductors with direct band gap of $\sim$ 1 eV, favoring their good performance in optoelectronic and photovoltaic fields. Subsequently, we have explored halogenated SL Ti$_{2}$O crystals, 1H- and 1T-Ti$_{2}$OX$_{2}$ (X= F, Cl and Br) monolayers. Their structure characteristics, stabilities and electronic properties are systematically explored and compared, and their relations are also scrutinized. The calculated absorptions of sunlight are from ultraviolet to near-infrared in the solar spectrum, suggesting that Ti$_{2}$OX$_{2}$ monolayers are promising donor materials for light harvesting in solar cells and infrared detector candidates. Besides, the excitons in 1H- and 1T-Ti$_{2}$OX$_{2}$ monolayers are relatively readily to be dissociated into free electrons and holes, along with considerably small binding energies (0.31 eV$-$0.51), favoring their outstanding photoelectric utilization efficiency. In combination with the \textit{ab initio} nonadiabatic molecular dynamics simulations, the carrier life of SL 1H- and 1T-Ti$_{2}$OX$_{2}$ are further evaluated. Our work wolud broaden the family of traditional MXene, and shed light on future rational design 2D MOene with multi-functions.


\begin{acknowledgement}
This work is supported by the Startup funds of Outstanding Talents of UESTC (A1098531023601205), National Youth Talents Plan of China (G05QNQR049), and the Open-Foundation of Key Laboratory of Laser Device Technology, China North Industries Group Corporation Limited (KLLDT202106) B.-T.W. acknowledge financial support from the Natural Science Foundation of China (Grants No. 11675195 and No. 12074381).
\end{acknowledgement}


\bibliography{paper}

\clearpage
\begin{figure*}[htbp!]
	\centering
	\includegraphics[width=\linewidth]{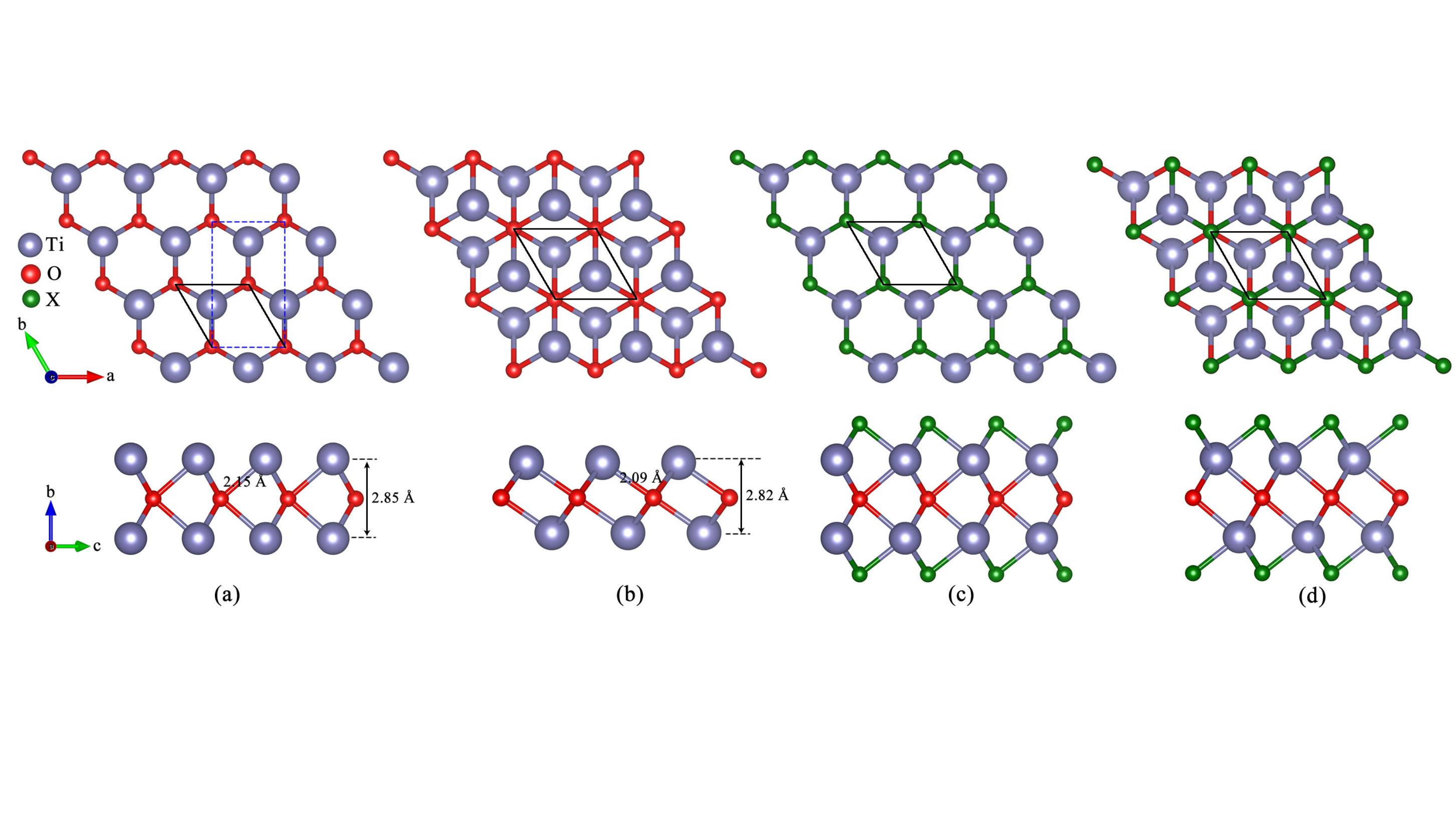}
	\caption{Top (upper panel) and side (lower panel) views for (a) 1H-Ti$_{2}$O, (b) 1T-Ti$_{2}$O in Ref. 52, (c) 1H- and (d) 1T-Ti$_{2}$OX$_{2}$ (X=F, Cl, Br and I) monolayer. The blank solid line represents the primitive cell, and blue dash line in (a) denotes conventional cell of SL 1H-Ti$_{2}$O.}
\end{figure*}

\begin{figure*}[htb!]
	\begin{center}
		\includegraphics[width=\linewidth]{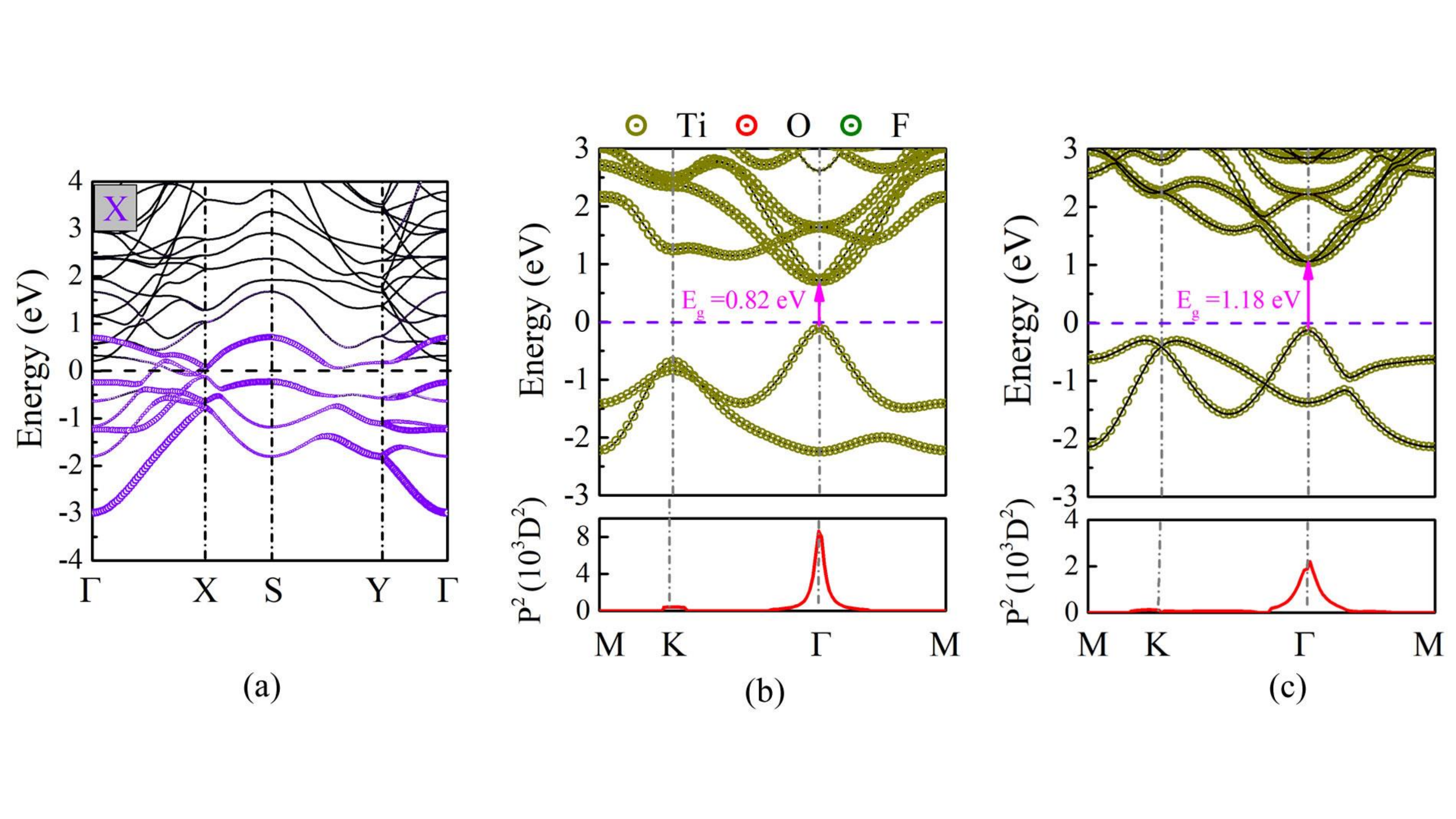}
	\end{center}
	\caption{(a) The projected spin-up electronic band structure weighted by anionic electrons, simulated by pseudo atoms (X aioms). The projected band structures magnituted by Ti, O and F orbitals (upper plane) and  calculated dipole transition dipole moment P$^{2}$ (lower plane) at the HSE06 level for SL (b) 1H- and (c) 1T-Ti$_{2}$OF$_{2}$. }
	\label{charge}
\end{figure*}

\begin{figure*}[htb!]
	\begin{center}
		\includegraphics[width=\linewidth]{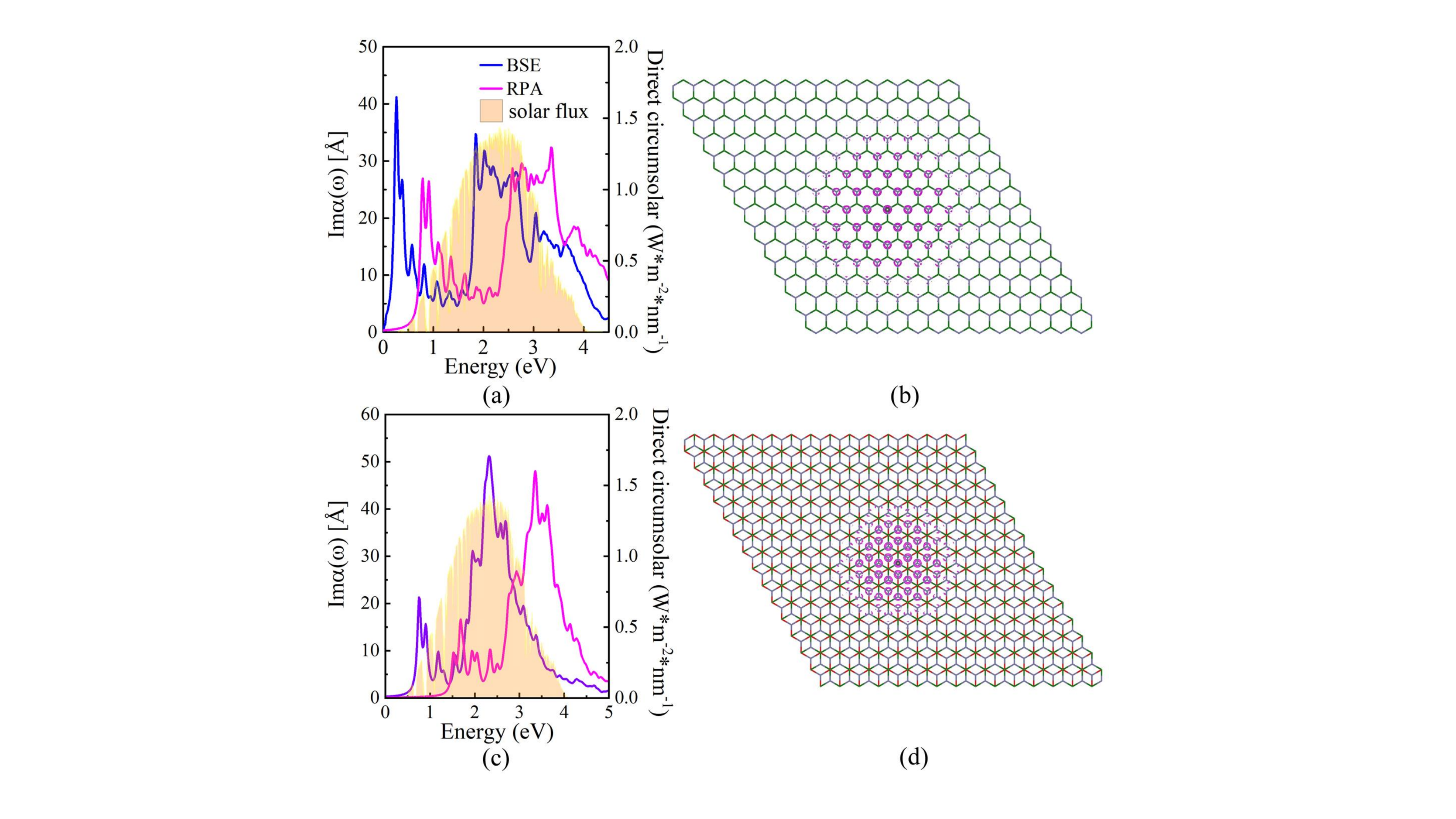}
	\end{center}
	\caption{The imaginary part of the transverse dielectric constant of (a) 1H- and (c) 1T-Ti$_{2}$OF$_{2}$ monolayers calculated with the RPA and BSE using the G$_{0}$W$_{0}$ quasiparticle correction, which are overlaped by standard AM1.5 G solar flux. Top side views of exciton wave function of (c) SL 1H- and (d) 1T-Ti$_{2}$OF$_{2}$. 15 $\times$ 15 unit cell is adopted and the hole position is marked by black spot in the center.}  
\end{figure*}

\begin{figure*}[htb!]
	\centering
	\includegraphics[width=\linewidth]{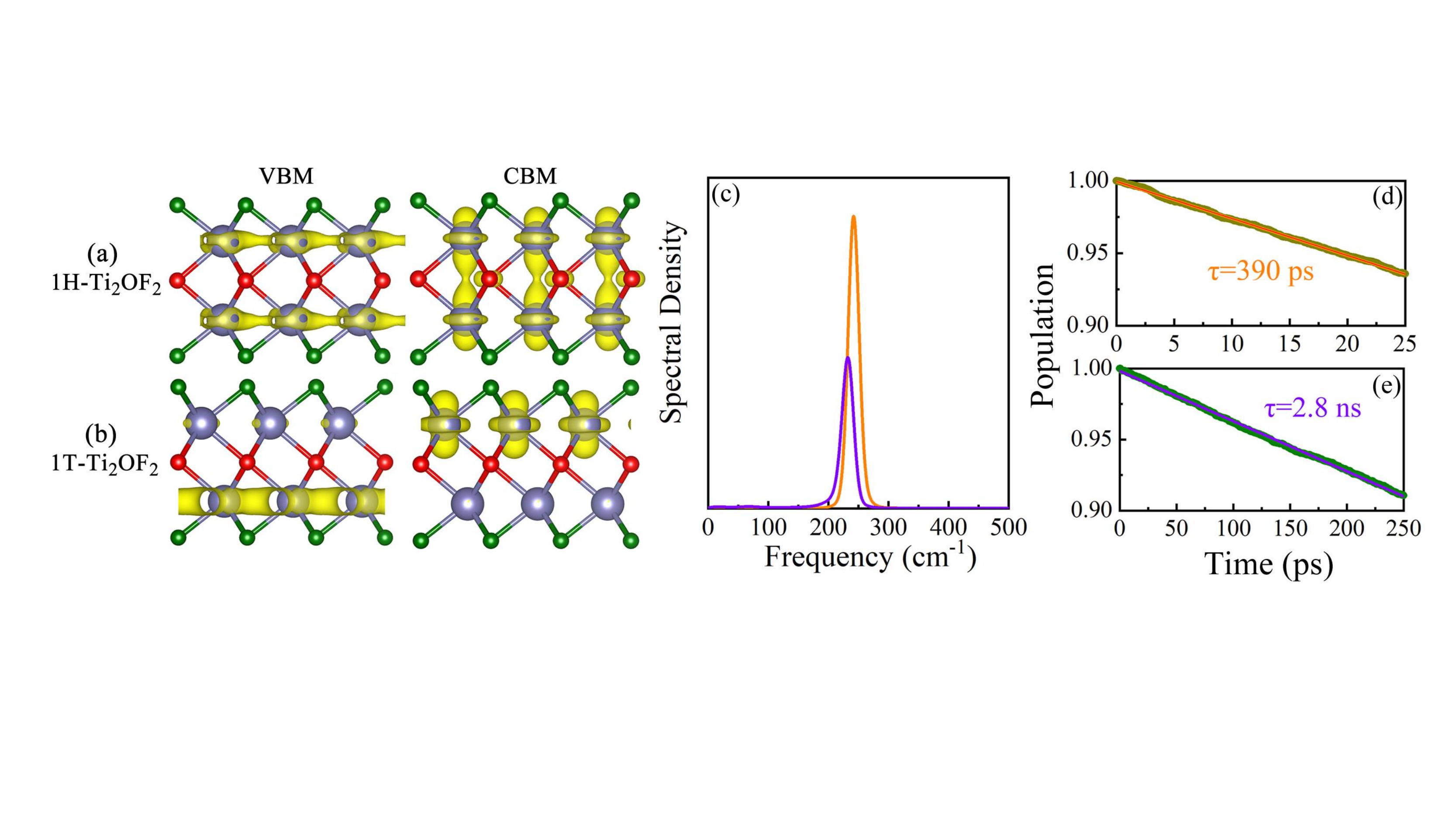}
	\caption{ Charge densities of VBM and CBM edge states for (a) 1H- and (b) 1T-Ti$_{2}$OF$_{2}$ monolayers. (c) The spectral desities,  electron-hole recombination dynamics for SL (d) 1H- and (e) 1T-Ti$_{2}$OF$_{2}$, simulated by excited-state population decay.}
\end{figure*}

\begin{table}[htbp]
	\small
	\caption{\ Corresponding relatively total energy (eV) to AFM-Néel spin configuration and magnetic moment ($\mu$B) of per Ti atom in SL 1H-Ti$_{2}$O.}
	\label{tbl:tbl-2}
	\begin{center}
		\begin{tabular}{cccccccccccccccc}
			\hline
			SL 1H-Ti$_{2}$O & Energy (eV) & Magnetic moment ($\mu$B)&\\
			\hline
			NM & 0.224 & 0 &\\
			FM & 0.071 & 0.57 &\\
			AFM-Zigzag & 0.208  & 0.23 &\\
			AFM-Néel & 0  & 0.59 &\\
			AFM-Stripy & 0.213 & 0.24 &\\
			\hline
		\end{tabular}
	\end{center}
\end{table}

\begin{table}[h]
	\small
	\caption{The energy of band gap E$_{g}$ at HSE06 level (eV), the optical band gap E$_{g}$$^{o}$ (eV), G$_{0}$W$_{0}$ gap (eV), and the excition binding energy E$_{b}$ (eV) for Ti$_{2}$OX$_{2}$ monolayers.}
	\label{tbl:example}
	\begin{center}
		\begin{tabular}{ccccccccccccccccccc}
			\hline
			Compounds&E$_{g}$&E$_{g}$$^{o}$&G$_{0}$W$_{0}$ gap&E$_{b}$\\
			\hline
			1H-Ti$_{2}$OF$_{2}$&0.82&0.28&0.72&0.44\\
			1H-Ti$_{2}$OCl$_{2}$&0.58&0.13&0.52&0.39\\ 
			1H-Ti$_{2}$OBr$_{2}$&0.07&0.16&0.49&0.33\\
			1T-Ti$_{2}$OF$_{2}$&1.18&0.76&1.37&0.61\\
			1T-Ti$_{2}$OCl$_{2}$&1.12&0.57&0.98&0.41\\
			1T-Ti$_{2}$OBr$_{2}$&0.92&0.57&0.88&0.31\\
			\hline
		\end{tabular}
	\end{center}
\end{table} 

\begin{table}[h]
	\small
	\caption{Absolute NA coupling (NAC), pure-dephasing time ($\tau$), Homogeneous line width (HLW), and NA relaxation time (T) for charge recombination in 1H- and 1T-Ti$_{2}$OF$_{2}$ at 300 K.}
	\label{tbl:example}
	\begin{center}
		\begin{tabular}{ccccccccccccccccccc}
			\hline
			Compounds&NAC (meV)&$\tau$ (fs)& HLW (meV)& T (ns)&\\
			\hline
			1H-Ti$_{2}$OF$_{2}$&3.10&11.26&58.53&0.39\\
			1T-Ti$_{2}$OF$_{2}$&2.46&20.11&32.76&2.80\\
			\hline
		\end{tabular}
	\end{center}
\end{table}

\end{document}